\documentclass[aps,prb,twocolumn,showpacs,superscriptaddress]{revtex4-1}
\usepackage{graphicx,amssymb,amsfonts,amsmath}

\usepackage{graphicx,bm}
\usepackage{epsfig}
\usepackage{color}
\usepackage{epsfig,amssymb,amsmath,amsthm,amsfonts,amsbsy,mathrsfs}
\usepackage{siunitx}

\usepackage{ulem}

\usepackage[utf8]{inputenc}

\usepackage{color}
\definecolor{darkblue}{rgb}{0,0,0.6}
\definecolor{darkred}{rgb}{0.6,0,0}
\definecolor{darkgreen}{rgb}{0,0.6,0}
\definecolor{purple}{rgb}{0.6,0,0.6}
\definecolor{gray}{rgb}{0.5,0.5,0.5}

\usepackage{soul}

\newcommand{\prom}[1]{\overline{#1}}

\newcommand{\AKcom}[1]{{{#1}}}

\DeclareSIUnit\oersted{Oe}

\begin{document}

\title{Intermittent collective dynamics of domain walls in the creep regime}

\author{Mat\'{\i}as Pablo Grassi}
\affiliation{Instituto Balseiro, Universidad Nacional de Cuyo - CNEA, Av. Bustillo 9500, 8400 S. C. de Bariloche, R\'{\i}o Negro, Argentina.}

\author{Alejandro B. Kolton}
\affiliation{Instituto Balseiro, Universidad Nacional de Cuyo - CNEA, Av. Bustillo 9500, 8400 S. C. de Bariloche, R\'{\i}o Negro, Argentina.}
\affiliation{CONICET, Centro At\'omico Bariloche, 8400 San Carlos de Bariloche, R\'{\i}o Negro, Argentina.}

\author{Vincent Jeudy}
\affiliation{Laboratoire de Physique des Solides, CNRS, Univ. Paris-Sud, Universit\'e Paris-Saclay, 91405 Orsay, France.}

\author{Alexandra Mougin}
\affiliation{Laboratoire de Physique des Solides, CNRS, Univ. Paris-Sud, Universit\'e Paris-Saclay, 91405 Orsay, France.}

\author{Sebastian Bustingorry}
\affiliation{Instituto de Nanociencia y Nanotecnología, CNEA--CONICET, Centro At\'omico Bariloche, 8400 San Carlos de Bariloche, R\'{\i}o Negro, Argentina.}

\author{Javier Curiale}
\affiliation{Instituto Balseiro, Universidad Nacional de Cuyo - CNEA, Av. Bustillo 9500, 8400 S. C. de Bariloche, R\'{\i}o Negro, Argentina.}
\affiliation{Instituto de Nanociencia y Nanotecnología, CNEA--CONICET, Centro At\'omico Bariloche, 8400 San Carlos de Bariloche, R\'{\i}o Negro, Argentina.}

\date{\today}

\begin{abstract}
{
We 
study the ultra slow domain wall motion in ferromagnetic thin films driven by a weak magnetic field. 
Using time resolved magneto-optical Kerr effect microscopy, we access to the
statistics of the intermittent thermally activated domain wall jumps 
between deep metastable states. 
Our observations are 
consistent with the 
existence of \textit{creep avalanches}: 
roughly independent clusters
with broad size and ignition waiting-time distributions, 
each one composed by a large number of spatio-temporally 
correlated thermally activated elementary 
events. Moreover, we 
evidence that the large scale geometry 
of domain walls 
is better described by depinning rather than equilibrium universal exponents.} 
\end{abstract}

\maketitle

\section{Introduction}
Domain walls (DW) in thin ferromagnetic films have become a paradigmatic system~\cite{lemerle_domainwall_creep, metaxas_depinning_thermal_rounding,Kim2009,Gorchon2014,Jeudy2016,DiazPardo2017} to learn about the universal interplay between disorder, elasticity and thermal fluctuations in driven interfaces. Such physics is relevant for a large variety of experimental 
systems~\cite{ledoussal_contact_line,Atis_prl2015,Chepizhko_pnas2016}, 
and for potential 
applications as DW are building blocks for proposed magnetic storage devices~\cite{Parkin11042008}. 
The caveat is that even an arbitrarily weak disorder has a rather dramatic effect on the DW dynamics, notably the occurrence of a depinning threshold~\cite{nattermann_creep_law,ioffe_creep}. Below the threshold, DW are pinned at zero temperature and they present a thermally activated glassy behavior called the creep regime at finite temperature.
A better understanding of the impact of disorder in the low velocity regimes is thus fundamental for a comprehensive study of DW dynamics, and of disordered elastic interfaces in general.

Most of the experimental studies on weakly driven 
DW motion, including very recent ones 
~\cite{metaxas_depinning_thermal_rounding,Kim2009,Gorchon2014,Jeudy2016,DiazPardo2017},
focused on the universal features of the steady 
DW mean velocity vs the field $H$ and temperature $T$, but not 
in its spatio-temporal fluctuations. 
{Such kind of study has been mostly performed close to the depinning threshold} 
{where the fluctuations are dominated by large deterministic collective events.}
For example, avalanche size distribution and its universal properties has been discussed in the context of Barkhausen noise \cite{Durin_prl2016}, contact lines of liquids \cite{ledoussal_contact_line}, crack propagation \cite{Laurson2013} and even in reaction fronts in disordered flows \cite{Atis_prl2015} and active cell migration \cite{Chepizhko_pnas2016}. {Well below the depinning threshold, the phenomenology of avalanches} 
{have remained much less clear}~\cite{Repain2004}. 
{Recently however, theoretical studies of ultra-slow creep 
motion~\cite{Ferrero2017} have unveiled rather unexpected and non trivial 
spatio-temporal patterns}, {whose elementary events 
strongly differ from those encountered close to the depinning threshold}.
Therefore, tackling experimentally a detailed statistical 
study of magnetization reversal events is particularly 
interesting.

{The numerical simulations reported in Ref.~\onlinecite{Ferrero2017} show that} creep motion 
of a one dimensional interface model proceeds 
via a sequence of \textit{elementary events} (EE) 
of fluctuating sizes. 
These EE are the minimal thermally activated 
jumps that make the DW 
overcome energy barriers and
irreversibly advance 
under the applied field $H$. 
The size statistics of EE display broad distributions,
with a characteristic lateral size  cut-off 
$L_{\tt opt} \sim H^{-3/4}$ and 
a characteristic area size $S_{\tt opt} \sim L_{\tt opt}^{5/3}$. 
These results confirm the existence of an optimal 
``thermal nucleous'', as proposed in the pioneer 
creep theories~\cite{nattermann_creep_law,ioffe_creep}.
Since energy barriers for DW motion scale as $U_{\tt opt} \sim L_{\tt opt}^{1/3} \sim H^{-1/4}$~(Ref.~\onlinecite{chauve_creep_long}), 
Arrhenius activation of these nuclei 
{leads} to the celebrated creep-law $\ln v \sim -H^{-1/4}/T$ 
for the mean velocity $v$ at which the DW move under the action of a small magnetic field $H$. 
The EE are not normally distributed in size and are not independent as traditionally assumed. On one hand,
below $S_{\tt opt}$, EE areas are power-law 
distributed as $P_{\tt EE}(S) \approx S^{-\tau_{\tt EE}}G(S/S_{\tt opt})$, 
with $\tau_{\tt EE}$ a characteristic exponent and 
$G(x)$ a rapidly decaying function for $x > 1$.
On the other hand, EE tend to cluster in space and time forming 
larger \textit{cluster events} (CE).
These CE are similar to the so called ``creep avalanches'' 
suggested by functional renormalization group calculations in 
Ref.~\onlinecite{chauve_creep_long} and experimentally noticed 
in Ref.~\onlinecite{Repain2004}.
Such composite objects are, unlike EE, weakly correlated and have 
a much broader distribution of areas, 
$P_{\tt CE}(S) \sim S^{-\tau_{\tt CE}}$ with $\tau_{\tt CE}$ 
a universal exponent.
{These interesting predictions were not yet 
evidenced experimentally nor confirmed} by other theoretical approaches.

In this \AKcom{work} we test the above scenario by a {statistical analysis of the ultra slow time evolution of magnetization reversal in ferromagnetic Pt/Co/Pt thin films. For different time windows of duration $\Delta t$, we determine the size ($S$) distribution $P_{\tt WE}(S) \equiv P_{\tt WE}(S;\Delta t,T,H)$ of the observed consecutive \textit{compact} magnetization reversal area that we call ``window-event'' (WE).
This procedure permits us to relate WE with EE and CE and to show that the features displayed by $P_{\tt WE}(S)$ are} 
{consistent with the picture summarized above of rare localized EE acting as epicenters of large CE or ``creep avalanches'', each made of a large number of spatio-temporally correlated EE. Furthermore, our analysis of the intermittent collective DW motion allows to characterize the statistics of waiting times between epicenter EE, thus going beyond the ``geometric'' predictions of Ref~\onlinecite{Ferrero2017}.}

\section{Methods}
Experiments were mainly performed on a Pt(4.5 nm)/Co(0.7 nm)/Pt(3.5 nm) thin ferromagnetic film with perpendicular magnetic anisotropy.
A polar magneto-optical Kerr effect (PMOKE) microscope was used to image magnetic domains. 
In order to characterize the DW dynamics, starting with a seed magnetic domain, a train of magnetic field pulses of duration $t$ and intensity $H$ were applied perpendicular to the film plane to favour the growth of the initial domain. The DW velocity was then computed following a standard differential protocol.
After identifying the creep regime in the $H-T$ plane by fitting the creep-law $\ln v \sim -H^{-1/4}/T$, we fix $T$ to two possible values, room temperature and $50~\si{\celsius}$, and choose $H=46.1~\si{\oersted}$ and $H=24.2~\si{\oersted}$ respectively, such that $v \sim 1~\si{\nano \meter \per \second}$ in each case.  
We then analyze the magnetization reversal events at each temperature, 
for a total applied field time $t=27000~\si{\second}$.
Since the characteristic areas of EE are 
expected to scale as $S_{\tt opt} \sim H^{-5/4}$, 
and the energy barriers for nucleation 
as $U_{\tt opt} \sim H^{-1/4}$, choosing 
fields deep in the creep regime
allows us to maximize, in principle, 
our spatial and temporal 
sensitivity to intrinsic collective events. 
For these fields we indeed observe a clear 
intermittent (i.e. not smooth) growth. 
To characterize it statistically, during the long-time magnetic 
field pulse 
we stroboscopically observe the growth at intervals $\Delta t$, such that $t \gg \Delta t$. 
The duration $\Delta t$ is much larger than the 
acquisition time of each image,
and much smaller than the
pulse time $t$ so to collect a large number of events. 
This allows us to compute
their area distribution, $P_{\tt WE}(S)$, 
for different $\Delta t$ and $T$. 
Although we mainly report results for one region of a specific sample, 
we have also performed less detailed but similar measurements in 
other regions of the same sample and also in a different material and checked 
robustness of our results. 
We discard WE touching any border of the region of interest in order to not 
underestimate their area and make a proper comparison with theoretical 
predictions. We have checked that this protocol does not affect the 
tails of $P_{\tt WE}(S)$ for the time windows $\Delta t$ used.
We refer the reader to appendix \ref{sec:samples} for further details 
on our experimental setup and protocols.
Magnetization reversal events were previously obtained 
in irradiated Pt/Co/Pt samples~\cite{Repain2004}, identifying 
between 30 and 50 events depending on field values. In the 
present work we were able to obtain thousands of 
WE, thus allowing a more precise statistical 
description, amenable to comparison with the universal 
theoretical predictions.

\section{Results}
\label{sec:results}
\subsection{Domain wall motion within the creep regime}

\label{sec:dwmotion}
\begin{figure}[!t]
\begin{center}
\includegraphics[width=0.9\columnwidth]{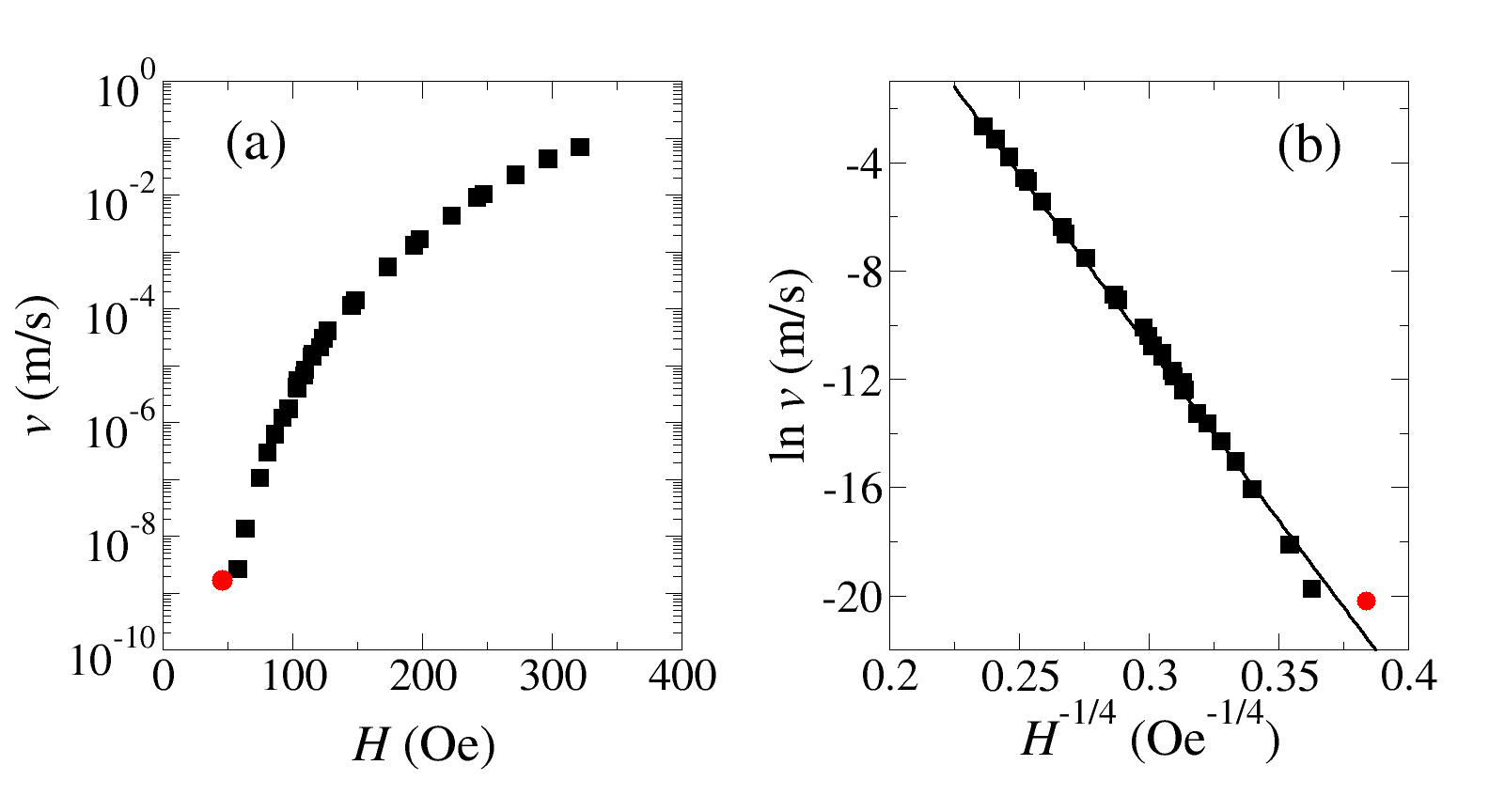}
\includegraphics[width=0.9\columnwidth]{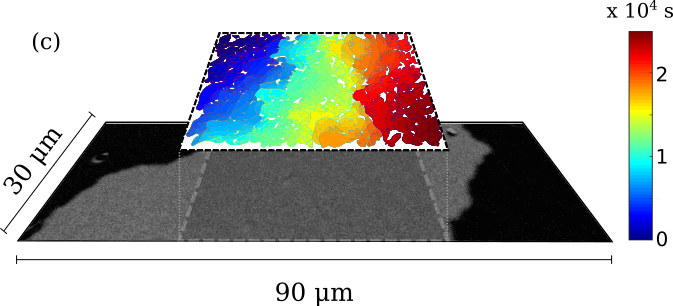}
\caption{(a)-(b) Velocity-field characteristics for ultrathin Pt/Co/Pt magnetic thin film in different scales. In (b) the fitted solid line confirms agreement with the creep velocity law: $\ln v \sim H^{-1/4}$. The red point corresponds to a single long pulse of duration $t=27000$~s at a small field of $H=46.1~\si{\oersted}$ and room temperature (RT). The total reversed area over this long pulse is indicated in (c) and corresponds to $v=1.7\,10^{-9}$~m/s. During this long pulse PMOKE images were taken every $t_0=15$~s, allowing to identify $N_{\tt WE}(t_0,t)=1151$ magnetization reversal events or ``window events'' (WE), highlighted over the image. The color scale corresponds to the time at which each WE was observed. 
}
\label{fig:creep-plot}
\end{center}
\end{figure}

The obtained field dependence of domain wall velocity for the analyzed sample is presented in 
Fig.~\ref{fig:creep-plot}(a). The figure shows the evolution of the velocity as 
a function of the magnetic field over eight orders of magnitude.
Within the creep regime, thermal activation over a field dependent energy barrier leads to a stretched exponential increase of the velocity, given by~\cite{nattermann_creep_law,ioffe_creep,lemerle_domainwall_creep, chauve_creep_long} 
\begin{equation}
 \label{eq:creep}
 v = v_0 \exp \left[ - \frac{T_d}{T} \left( \frac{H}{H_d} \right)^{-\mu} \right],
\end{equation}
where $v_0$ is a temperature dependent velocity~\cite{Gorchon2014}, $T$ the temperature, $k_B T_d$ a typical energy scale coming from the competition between elasticiy and disorder ($k_B$ being the Boltzmann constant), $H_d$ the depinning field and $\mu=1/4$ the universal creep exponent. 
As shown in Fig.~\ref{fig:creep-plot}(b), a straight line with a negative slope 
in a plot of $\ln v$ against $H^{-1/4}$ confirms that the measured velocities are within the creep regime, and in addition that the system belongs to the universality classes of one dimensional elastic systems displacing in a two dimensional media, with a random-bond type of disorder and short-range elasticity. 
The fit to the creep formula of Eq.~(\ref{eq:creep}) for the two temperatures we analyzed are 
\begin{eqnarray}
\ln[v ~\si{\meter}^{-1} \si{\second}] &=& -128(1)(\si{\oersted})^{1/4} H^{-1/4} + 27.6(3), 
\label{eq:vRT} \\
&=& -100(2)(\si{\oersted})^{1/4} H^{-1/4} + 24.4(5)
\label{eq:v50C}
\end{eqnarray}
at $T=$~RT [Eq.(\ref{eq:vRT})] and $T=50\si{\celsius}$  [Eq.(\ref{eq:v50C})].
This data, the experimental estimates for the depinning field $H_d$, the depinning temperature $T_d$ and key characteristic scales are reported in Table~\ref{tabla}.
%
\begin{table}[h]
\centering
\caption{Characteristic depinning values for the studied temperature and field values: $H_d$ is the depinning field, $T_d$ is the depinning temperature, $H_d^{1/4}\frac{T_d}{T}$ is the slope of the creep plot [see Eqs.~(\ref{eq:vRT}) and (\ref{eq:v50C})], and $T/T_d$ and $\left(\frac{T}{T_d}\right)\left(\frac{H}{H_d}\right)^{1/4}$ are related to the distribution of waiting times as discussed in Sec.~\ref{sec:waiting-times}.}
\label{tabla}
\begin{tabular}{|l|l|l|}
\hline
$T[\si{\kelvin}]$ & 293         & 323        \\ \hline
$H[\si{\oersted}]$ & 46.1        & 24.2        \\ \hline
$T_d[\si{\kelvin}]$ & 7142        & 6369       \\ \hline
$H_d[\si{\oersted}]$ & 760         & 650        \\ \hline
$H_d^{1/4}\frac{T_d}{T}[\si{\oersted^{1/4}}]$ & 128     & 100       \\ \hline
$\frac{T}{T_d}$ & 0.04    & 0.06    \\ \hline
$ \left(\frac{T}{T_d}\right)\left(\frac{H}{H_d}\right)^{1/4}$ & 0.02    & 0.02  \\ \hline
\end{tabular}
\end{table}

With the aim of pursuing the characterization of small magnetization reversal events responsible of the creep motion of elastic systems, one should consider that the typical area size $S_{\tt opt}$ of the 
``optimal thermal nuclei'' responsible for the velocity of Eq.(\ref{eq:creep}) 
dramatically increases when the magnetic field decreases, as $S_{\tt opt} \propto H^{-\nu_{\tt eq}(1+\zeta_{\tt eq})}$,
with $\nu_{\tt eq}$ and $\zeta_{\tt eq}$ positive universal exponents
~\cite{nattermann_creep_law,ioffe_creep,lemerle_domainwall_creep, chauve_creep_long}. Since velocity follows an streched exponential dependence with $S_{\tt opt}$ ~\cite{nattermann_creep_law,ioffe_creep}, decreasing the magnetic field implies to perform very long time experiments. Therefore, after nucleation of a single domain, a small magnetic field ($H=46.1$~Oe) is applied during a single long time pulse ($t=27000~\si{\second}$, i.e. $7.5$ hours), reaching a velocity $v=1.7\,10^{-9}$~m/s.  The velocity-field data thus obtained is indicated as a red point in Fig.~\ref{fig:creep-plot}(b), and the differential image shown in Fig.~\ref{fig:creep-plot}(c) corresponds to the full displacement of the domain wall under these conditions.
In order to identify magnetically reversed regions (WE), during the total long pulse time $t$, PMOKE images were taken every $t_0=15~\si{\second}$, which corresponds to the minimum time window $\Delta t$. During the image acquisition the magnetic field remained always ON.
Consecutive images were subtracted and, since the velocity is small ($v=1.7\,10^{-9}$~m/s), most of differential images do not show a clear advance of the domain wall. Eventually a magnetization reversal region resulting in a local advance of the domain wall position is observed. 
After the whole long time pulse experiment, the total reversed area (indicated in Fig.~\ref{fig:creep-plot}(c)) is fragmented into many small spatially compact regions obtained from the subtraction of consecutive images taken after $t_0$. The number of WE, is $N_{\tt WE}(t_0,t) = 1151$ and are highlighted with a color code over the image of the reversed area in Fig.~\ref{fig:creep-plot}(c).
Due to the characteristics of the used PMOKE microscope and the image analysis, the smallest detectable displacement of the domain wall correspond to events close to 0.3~$\si{\micro \meter^2}$ (25 pixels).

\subsection{Event areas}

\begin{figure}[!ht]
\begin{center}
\includegraphics[width=0.9\columnwidth]{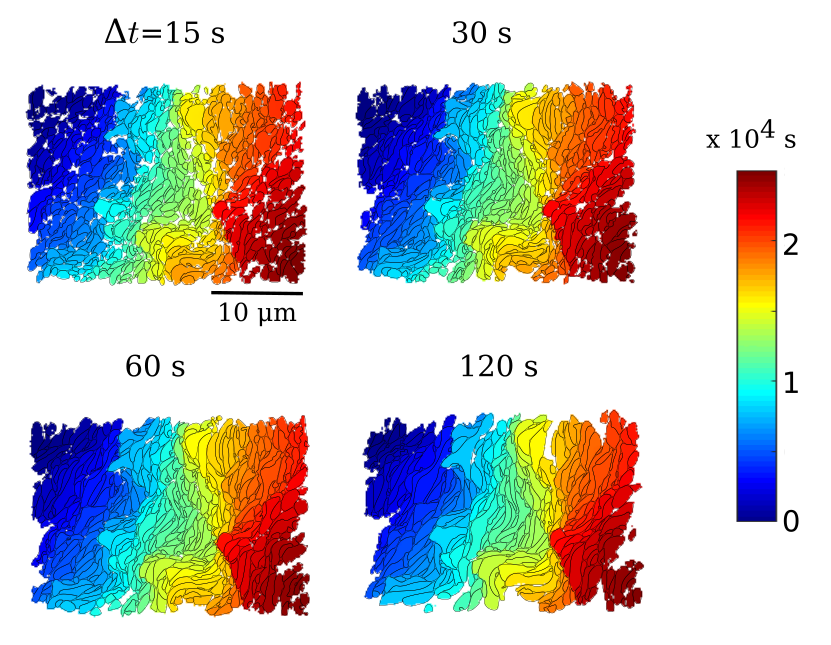}
\caption{Sequences of magnetization reversal areas (WE) detected 
for different time windows of duration $\Delta t$, for $T=$~RT and $H=46.1~\si{\oersted}$. 
{The color scale corresponds to the time at which each WE, delimited by contours lines, was detected.}}
\label{fig:coalescence}
\end{center}
\end{figure}    
In Fig.~\ref{fig:coalescence} we show typical WE sequences, for four different values of $\Delta t$, from a $15~\si{\second}$ to $120~\si{\second}$. 
We can appreciate that, for a given growth, each $\Delta t$ induces a particular partition of the total reversed area of the sequence.
At large $\Delta t$  the coalescence of several smaller WE corresponding to smaller $\Delta t$ becomes evident.

\begin{figure}[!ht]
\begin{center}
\includegraphics[width=1\columnwidth]{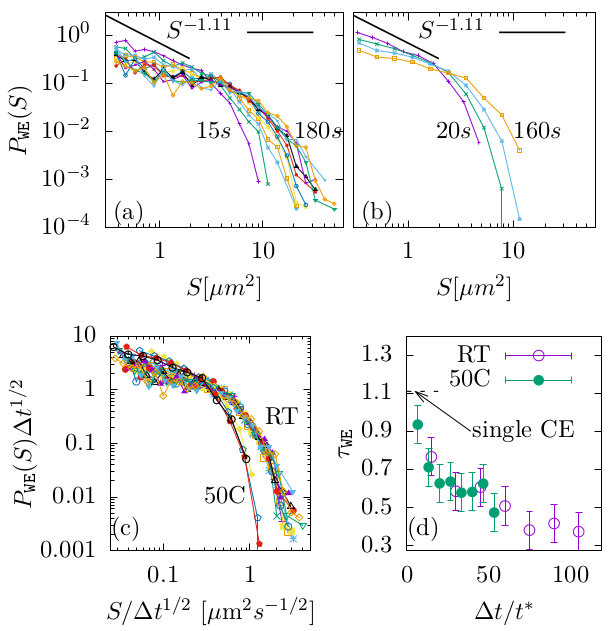}
\caption{WE area distributions for increasing window times $\Delta t$ (as indicated) at RT and $H=46.1~\si{\oersted}$ (a) and at $T=50~\si{\celsius}$ and $H=24.2~\si{\oersted}$ (b). In both cases $v\sim 1~\si{\nano \meter \per \second}$. At small $S$ we compare the initial decay of $P_{\tt WE}(S)$ with $S^{-\tau_{\tt CE}}$, 
with $\tau_{\tt CE} \approx 1.11$, where $\tau_{\tt CE}$ corresponds to 
depinning avalanches. 
(c) The collapse scaling shows that the data of (a) and (b) displays a large size cut-off scaling $S_{\tt WE}\sim (\Delta t/t^*)^{1/2}$, with $t^*$ an $H$ and $T$ dependent characteristic time.
(d) Effective power-law exponents $\tau_{\tt WE}$ for $P_{\tt WE}(S)$ vs $\Delta t/t^*$.
}
\label{fig:hist}
\end{center}
\end{figure}

In Fig.~ \ref{fig:hist}(a),(b) we compare 
size distributions $P_{\tt WE}(S)$, from 
$\Delta t=15$ to $180~\si{\second}$ at room temperature 
$T=$~RT and a field $H=46.1~\si{\oersted}$, and from 
$\Delta t=20$ to $160~\si{\second}$
at $T=50~\si{\celsius}$ and a field $H=24.2~\si{\oersted}$, respectively. 
The first remarkable 
feature of all these 
distributions is their broadness, which can be roughly 
described by 
$P_{\tt WE}(S) = S^{-\tau_{\tt WE}} G_{\tt WE}(S/S_{\tt WE})$, 
where $\tau_{\tt WE}$ 
is an effective power-law exponent and $S_{\tt WE}$ 
the cut-off value such that the function 
$G_{\tt WE}(x)$ is constant for small $x$ and 
decays faster than a power-law for $x \gtrsim 1$. 
{Quantitatively similar size-distributions were observed 
in different regions of the same sample and also in other 
kind of magnetic films (see appendix \ref{sec:samples}).

Both $\tau_{\tt WE}$ and the large-size 
cut-off $S_{\tt WE}$ depend on 
$\Delta t$.  
As can be appreciated in Fig.~\ref{fig:hist} (a)-(b)
$S_{\tt WE}$ increases with $\Delta t$, more 
specifically  
$S_{\tt WE} \sim (\Delta t/t^*)^{1/2}$. 
The fair collapse of $P(S)\Delta t^{1/2}$ vs 
$S/\Delta t^{1/2}$ shown in Fig.~\ref{fig:hist}(c)
confirms this dependence. 
Here, $t^*\equiv t^*(T,H)$ is a characteristic time. 
Concomitantely, in Fig.~\ref{fig:hist}(d) we show that $\tau_{\tt WE} \approx 1$ for the smallest $\Delta t/t^*$ for the whole data of Figs \ref{fig:hist}(a),(b). Note also that the same $t^*$ that describes the $S_{\tt WE}$ $(T,H)$-dependence allows to build a master curve for $\tau_{\tt WE}$ vs. $\Delta t/t^*$. For the characteristic times $t^*$ we find $t^*_{50C} \approx 1~\si{\second}$ at $T=50~\si{\celsius},~H=24.2~\si{\oersted}$ and $t^*_{RT} \approx t^*_{50C}/3$ at $T=$~RT, $H=46.1~\si{\oersted}$.
Therefore $S_{\tt WE}\approx (\Delta t)^{1/2}~\si{\micro \meter \squared \second^{-1/2}}$ in the first case, and 
$S_{\tt WE} \approx (3\Delta t)^{1/2}~\si{\micro \meter \squared \second^{-1/2}}$ in the second one.

Since EE of Ref.~\onlinecite{Ferrero2017} are 
power-law distributed with an exponent 
$\tau_{\tt EE}\approx 1.17$ it is tempting to directly compare small $\Delta t$ WE, 
which are also typically small, to EE.
A rough estimate for the Pt/Co/Pt films we study 
shows that the largest EE are of the order of 
$S_{\tt opt} = 10^{-3} (H_d/H)^{1.25} ~\si{\micro \meter \squared}$, 
where $H_d$ is the depinning field~\footnote{See Supp. Mat. in Ref.\onlinecite{Ferrero2017}.}.
Since $H_d \approx 637~\si{\oersted}$, and our lowest 
field is $H=46~\si{\oersted}$, we get that 
$S_{\tt opt} \sim 10^{-5} ~\si{\micro \meter \squared}$, which 
is clearly well below our PMOKE resolution 
of roughly $0.3~\si{\micro \meter \squared}$ ($25$ pixels).
We thus conclude that our detected WE can not be single EE, 
but the sum of a large number of them.
Namely, if in a time window $\Delta t$ we have 
${\cal N}_{\tt EE}$ 
such events, of sizes ${s}_1, {s}_2, ..., {s}_{{\cal N}_{\tt EE}}$, 
compactly grouped in a WE, its random area is 
${\cal S}_{\tt WE} \approx \sum_{i=1}^{{\cal N}_{\tt EE}} {s}_i$. 
The statistics of ${\cal S}_{\tt WE}$ thus directly relates 
to the statistics of EE random sizes $s_i$ contributing to the same WE
and of their $\Delta t$ dependent and fluctuating number 
${\cal N}_{\tt EE}$.

Given the small area of the EE  
compared to our detected WE, 
a pure 
statistical analysis is convenient.
If the EE were considered independent and accumulating 
at a well defined rate on each WE, 
by virtue of the central limit theorem 
we would naively expect $P_{\tt WE}(S)$ 
to develop an approximate gaussian shape around 
$\prom{{\cal N}_{\tt EE}} \prom{s}$.
$P_{\tt WE}(S)$ shows no 
tendency to approximate a normal 
nor even a peaked distribution
however: it is broad, 
even for $\Delta t$ in the minutes time scale.
To interpret this it is worth 
recalling that the central limit theorem tell us that
${\cal S}_{\tt WE} \approx \sum_{i=1}^{{\cal N}_{\tt EE}} s_i$ 
should converge to a Gaussian distribution if 
${\cal N}_{\tt EE}$ is large enough and the $s_i$ have 
finite variance and short-ranged correlations~\cite{Bouchaud1990}.
The EE have finite variance and, although they appear to be spatially correlated, 
there is no evidence of correlation between their areas~\cite{Ferrero2017}.
We hence interpret that ${\cal N}_{\tt EE}$ must be a strongly 
fluctuating quantity for all the 
$\Delta t$ analysed. 
Indeed, we experimentally observe for a fixed $\Delta t$ well defined bursts of magnetic activity, 
with ${\cal S}_{\tt WE} \gg 0.3~\si{\micro \meter \squared}$, 
coexisting with WE in the resolution edge 
${\cal S}_{\tt WE} \gtrsim 0.3~\si{\micro \meter \squared}$, at the same $H$ and $T$. 
Since any PMOKE
resolved  
area ${\cal S}_{\tt WE} > 0.3~\si{\micro \meter \squared}$ has
a large number of EE we arrive to the 
first important {observation} of our paper:
~{EE are strongly clustered spatio-temporally}.

\subsection{Domain Wall roughness}

The \AKcom{results of the previous section} are consistent with the EE clustering predicted 
for simple domain wall models~\cite{Ferrero2017,Purrello2017}. 
To go beyond, since EE are too small to be experimentally resolved, 
one is inmediately tempted to compare our experimentally resolved WE with 
the predicted CE. 
Indeed, unlike EE, CE are not expected 
~\cite{Ferrero2017} to be strongly 
correlated as we also observe for WE.
Moreover, the predicted value for $\tau_{\tt CE}\approx 1.11$ is only 
slightly above $\tau_{\tt WE}\approx 1$ 
observed in Fig.~\ref{fig:hist}(d) 
for the smallest $\Delta t$.
To argue that WE may indeed approach the single 
intrinsic 
CE  
in the small $\Delta t$ limit, 
we start by noting that 
the same scaling of zero temperature 
depinning avalanches,
$S_i \sim L_i^{1+\zeta_d}$, 
is also expected for CE~\cite{Ferrero2017} at finite 
temperature. 
In Fig.~\ref{fig:expo}(a) we analyse 
for $T=$~RT the approximately oblong shapes of WE 
by plotting the areas 
$S_i$ of each WE 
sampled from a long sequence, versus their 
corresponding lateral size $L_i$, 
defined as the major axis length of the reversed blobs. 
A crossover 
is observed at $S\approx 2 ~\si{\micro \meter \squared}$ 
below which we observe a $S_i \sim L_i^{2.25}$ 
scaling
\footnote{We show both the original 
noisy $S_i$ vs. $L_i$ curve and a smoothed 
curve, obtained by grouping similar areas in 
small bins and by 
assigning the average of $L_i$ to each area bin. 
Indistinguishable results are obtained by binning $L_i$ 
rather than $S_i$.}. 
{
The two main canditate depinning universality 
classes that are consistent with the observed 
creep law 
$\ln v \sim H^{-1/4}/T$
are the 1d quenched-Edwards-Wilkinson (qEW), 
and the 1d quenched-Kardar-Parisi-Zhang (qKPZ).
The first predicts $\zeta_d \approx 1.25$~\cite{FeBuKo2013}
while the second $\zeta_d \approx 0.63$~\cite{rosso_dep_exponent,tang_anisotropic_depinning,Barabasi-Stanley}.
} 
{Only} the qEW value is in good quantitative 
agreenment with Fig.~\ref{fig:expo}(a), in the 
small size WE limit~\footnote{The downward deviation 
crossover for large WEs or lateral 
acceleration is due to coalescence effects 
in the proximity of a percolation transition. 
See Sec.\ref{sec:discusions} 
for further details.}. In addition  
Fig.~\ref{fig:hist}(d) is quantitatively consistent with the 
relation $\tau_{\tt CE}=2-2/(1+\zeta_d)\approx 1.11$  
predicted for qEW. 
To investigate this issue in Fig.~\ref{fig:expo}(b) 
we computed the squared width 
{$W^2(L)\equiv \overline{u^2_L(x)}-\overline{u_L(x)}^2$
from different small segments of size $L$ extracted 
from typical DW configurations, where 
$u_L(x)$ is the DW displacement measured with 
respect to the untilted segment} (see Appendix \ref{sec:roughnessmethod} details).
The scaling $W^2\sim L^{2\zeta_d}$ is 
consistent {with the qEW \it{depinning roughness exponent}} $\zeta_d=1.25$ and thus with Figs. ~\ref{fig:hist}(d) 
and~\ref{fig:expo}(a). 
We then arrive to the 
second important observation of our paper: 
WE approach single 
CE in the small 
$\Delta t$ limit and we {find experimental evidence 
that 
the DW roughness and CE statistical properties} are better described by depinning rather 
than equilibrium exponents~
{ as {theoretically} 
predicted in Refs.~\onlinecite{chauve_creep_long,kolton_depinning_zerot2,kolton_dep_zeroT_long}}.
{
A smaller roughness exponent  
$0.69\pm 0.07$ was observed in extended DW in the 
same material in the pionering work by Lemerle {\it et. al.}
~\cite{lemerle_domainwall_creep}, and interpreted 
to be the equilibrium exponent $\zeta_{\tt eq} = 2/3$. 
Such interpretation implies an observation scale below
$L_{\tt opt}$~\cite{kolton_depinning_zerot2}.
However, from Ref.~\onlinecite{lemerle_domainwall_creep}
we infer $L_{\tt opt}\approx 0.18~\mu m$, lower 
than their PMOKE resolution of $0.28~\mu m$.
We thus conclude that a spatial crossover from the qEW value 
$\zeta_d \approx 1.25$ to a {\it non-equilibrium} 
exponent $\sim 0.69 \pm 0.07$ must exist.
A natural candidate is the Quenched Kardar-Parisi-Zhang (qKPZ) or
directed percolation depinning exponent 
$\zeta_d \approx 0.63$.
}
\begin{figure}[!ht]
\begin{center}
\includegraphics[width=\columnwidth]{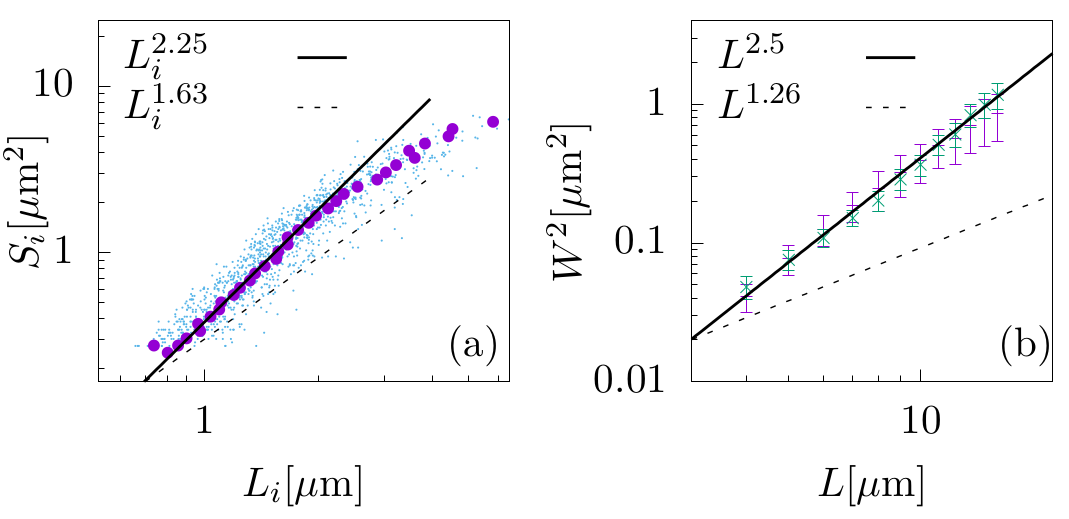}
\caption{
(a) Aspect ratio scaling of $\Delta t=15~\si{\second}$ WE. The solid (dashed) 
line shows the expected 
depinning scaling $S_i \sim L_i^{1+\zeta_d}$ for qEW (qKPZ) class.
(b) Scaling of the square width $W^2$ of DW segments of size $L$, for two typical configurations at RT.
The solid (dashed) line shows the expected qEW (qKPZ) scaling at depinning, $W^2 \sim L^{2\zeta_d}$, with 
$\zeta_d=1.25 (0.63)$.
}
\label{fig:expo}
\end{center}
\end{figure}

\subsection{Event lengths}
In Fig.~\ref{fig:pofl}(a) we show the areas $S_i$ vs the major axis lenght $L_i$
of each WE. \AKcom{The difference with Fig.~\ref{fig:expo}(a), where $\Delta t=15s$, 
is that now we plot WE for all the $\Delta t$, from $15\si{\second}$ to $180\si{\second}$, 
in order to observe the effects of large WE.}
We show both the cloud obtained from raw data and an 
averaged version by grouping areas in small logarithmically increasing 
bins and by taking the corresponding average value of $L_i$ in such groups.
We compare with 
the depinning scaling $S_i \sim L_i^{1+\zeta_d}$, expected 
for cluster events (CE) in the creep regime~\cite{Ferrero2017}, both 
for the qEW class 
where $\zeta_d=1.25$ and for the 
qKPZ class, 
where $\zeta_d\approx 0.63$. 
At $L_i \approx L^* = 2\si{\mu \meter}$ a clear 
crossover is observed (indicated by the vertical line). 
As can be appreciated in the figure, for $L_i<L^*$ 
a better agreement is obtained for qEW, 
as compared for instance with the qKPZ class.

In Fig.~\ref{fig:pofl}(b) we show the 
(non normalized) probability distribution $P_{\tt WE}(L)$ 
for all the $L_i$ observed. As for $P_{\tt WE}(S)$, 
we observe a broad distribution. 
If WE were a part, single, or dominated by single 
CE we expect indeed WE to display 
power law distributions similar to the 
ones observed for depinning avalanches,
$P_{\tt CE}(S) \sim S^{-\tau_{\tt CE}}$ and 
$P_{\tt CE}(L) \sim S^{-\tau_L}$, where $\tau_{\tt CE}$ and $\tau_L$ are related to depinning exponents.
The general exponents are well known~\cite{zapperi1998} 
\begin{eqnarray}
\tau_{\tt CE} &=& 2 - (\zeta_d+1/\nu_d)/(1+\zeta_d),\label{eq:tauS}\\
\tau_L &=& \tau_{\tt CE} (1+\zeta_d) - \zeta_d \label{eq:tauL}
\end{eqnarray}
where $\zeta_d$ is the depinning 
roughness exponent and $\nu_d$ the depinning correlation length exponent. 
For the qEW universality class, we have $\zeta_d\approx 1.25$ 
and, by virtue of the statistical 
tilt symmetry~\cite{kardar_review_lines}, $\nu_d=1/(2-\zeta_d)\approx 1.33$. 
On the other hand $\zeta_d\approx 0.63$ and 
$\nu_d\approx 1.73$ for the qKPZ class~\cite{Barabasi-Stanley}, where 
the statistical tilt symmetry is broken.
This yields $\tau_{\tt CE} \approx 1.11$, $\tau_L\approx 1.25$ 
for the qEW class, and 
$\tau_{\tt CE} \approx 1.25$, $\tau_L\approx 1.42$ for the qKPZ 
class. The effective power law at intermediate 
$L\lesssim L^*$ (indicated by the vertical line) 
is roughly consistent with qEW. Unfortunately however, 
the effective power law observed  
in Fig.~\ref{fig:pofl}(b) is roughly consistent with both 
classes, unlike Fig.~\ref{fig:pofl}(a) which is more 
consistent with the qEW class.

The crossover at $L^*$, observed in Figs.~\ref{fig:pofl}(a), 
may be associated to the CE coalescence process occuring 
for large WE~\footnote{In our protocol, 
the maximum WE lateral size $L_i$ was limited 
by the lateral size $L_{\tt roi}$ of the region of interest. 
We did not consider WE touching nor spanning 
completely the region of interest. If spanning WE were considered 
we would have $S_{\tt WE} \approx v \Delta t L_{\tt roi}$ in the long-time limit, 
with $v$ the average velocity, shown in Figs.~\ref{fig:creep-plot}(a),(b).}. 
In that case, a WE area can be written 
as a sum of a given number ${\cal N}_{\tt CE}$ of CE areas.  
${\cal S}_{\tt WE} \approx \sum_{i=1}^{{\cal N}_{\tt CE}} {\cal S}_i$.
Since $P_{\tt CE}(S)$ is a broad distribution, 
the typical WE area is dominated by the largest areas and thus 
relates to the typical number of 
CE as $S_{\tt WE} \sim N_{\tt CE}^{1/(\tau-1)}$. 
The lateral size of a WE instead satisfies
an inequality $L_{\tt WE} < \sum_{i=1}^{{\cal N}_{\tt CE}} {\cal L}_i$, 
as the fluctuating CE lateral extensions ${\cal L}_i$ can now overlap. 
Since $P_{\tt CE}(L)$ is also a broad distribution, 
we can use the same extreme value argument to estimate
$L_{\tt WE} \lesssim N_{\tt CE}^{1/(\tau_L-1)}$. Combining these 
results we get $S_{\tt WE} \gtrsim L_{\tt WE}^{(\tau_L-1)/(\tau-1)} \equiv L_{\tt WE}^{1+\zeta_d}$.
This shows that WE areas should scale with their length approximately as CE in the $L_i < L^*$ regime, 
as observed in Fig.~\ref{fig:pofl}(a). 
Above $L^*$ however, where large WE become a non negligible fraction of the interface, 
the last scaling prediction breaks down. In section \ref{sec:discusions}
we discuss a simple model that quantitatively 
accounts for the crossover observed at $L^*$ 
in the $S_i$ vs $L_i$ plot.


\begin{figure}[!ht]
\begin{center}
\includegraphics[width=0.9\columnwidth]{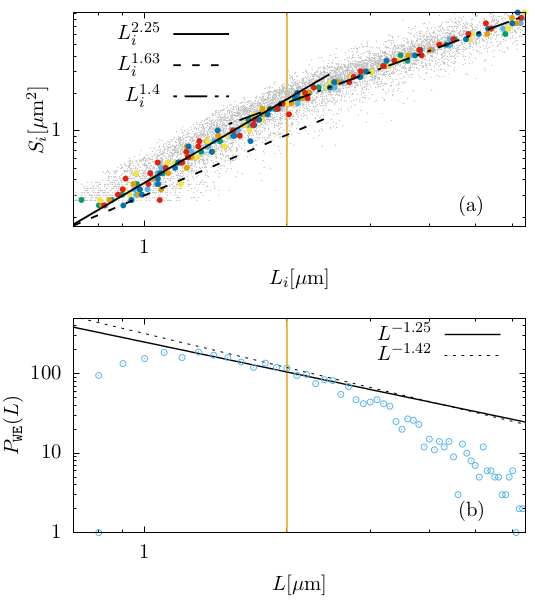}
\caption{(a) Aspect-ratio of WEs obtained experimentally, between 
the area $S_i$ and major axis length $L_i$ of individual WE. 
We 
display the data $S_i$ vs $L_i$ for WE corresponding to 
all values of $\Delta t$, in order to enhance the crossover 
behaviour and access large WE. Small WE fairly scale as CE or 
depinning avalanches in the qEW class, $S_i \sim L_i^{1+\zeta_d}$ 
with $\zeta_d \approx 1.25$ and thus $S_i \sim L_i^{2.25}$ (solid line), 
as compared to the qKPZ class, with $\zeta_d=0.63$, predicting 
$S_i \sim L_i^{1.63}$ (dashed line). 
The dotted-dashed line indicates a fair 
$S_i \sim L_i^{1.4}$ scaling, which can be rationalized 
using a simple model. (b) Probability distribution (non normalized) 
for the major axis length $L_i$ of the events shown in (a). 
In (a) and (b) a vertical line indicates the approximate location
of the aspect-ratio crossover, $L^*$.
}
\label{fig:pofl}
\end{center}
\end{figure}

\subsection{Waiting times}
\label{sec:waiting-times}

The behaviour at large $\Delta t$, where 
the probability to observe single CE in a WE decreases, 
is directly related to the behaviour of the large-size $P_{\tt WE}(S)$ cut-off, $S_{\tt WE}$, with $\Delta t$.
In such 
regime we can regard each WE area as the sum of a given number ${\cal N}_{\tt CE}$ of cluster areas, ${\cal S}_{\tt WE}=\sum_{i=1}^{{\cal N}_{\tt CE}} {\cal S}_i$.
As ${\cal N}_{\tt CE}$ can only grow irreversibly with $\Delta t$, so does the large size cut-off $S_{\tt WE}$. Naively one may think that $S_{\tt WE}$ should linearly increase with $\Delta t$ because the sum of all WE areas observed in a region of a fixed lateral size $L$ should grow as $L v \Delta t$ in a steady-state regime.
As shown in Fig.~\ref{fig:hist}(c) we find instead a sub-linear increase $S_{\tt WE} \sim (\Delta t/t^*)^{1/2}$.
To make sense of this striking observation it is instructive to regard the area ${\cal S}_{\tt WE}$ vs. $\Delta t$ as a continous-time continous-jump random-walk, with random CE area increments 
${\cal S}_i$ and 
waiting times $\delta_i$ for the ignition of a new CE, 
such that $\Delta t = \sum_{i=1}^{{\cal N}_{\tt CE}} \delta_i$.
If we assume that
the $\delta_i$ are distributed according to 
$\psi(\delta) \sim  {t^*}^{\alpha} \;\delta^{-(1+\alpha)}$, 
with $0< \alpha \leq 1$ 
{characterizing the broadness of $\psi(\delta)$}, 
we get $\Delta t \sim t^* N_{\tt CE}^{1/\alpha}$ 
for the typical number of events $N_{\tt CE}$ in a $\Delta t$. 
Since the same heuristic arguments apply for the broadly distributed CE we get 
$S_{\tt WE} \sim N_{\tt CE}^{1/(\tau_{\tt CE}-1)}$. Combining the two last results 
we get $S_{\tt WE} \sim (\Delta t/t^*)^{\alpha/(\tau_{\tt CE}-1)}$, 
which fairly describes our data of Fig.~\ref{fig:hist}(c) if $\alpha/(\tau_{\tt CE}-1) \approx 1/2$. 
Using $\tau_{\tt CE} \approx 1.11$ we obtain $\alpha \approx 0.05$.

{
Broad waiting time distributions 
have been heuristically
derived for creep motion~\cite{chen_marchetti_vinokur_1996, monthus_garel_2008}, 
borrowing ideas from more general random energy models 
(see for instance Ref.~\onlinecite{Bouchaud1990}),
and also observed numerically close to the depinning threshold~\footnote{E. A. Jagla, unpublished.}. 
The basic idea 
is to assume that the 
EE barrier distribution 
behaves as
$P(U)\sim \exp[-U/U^*]/U^*$ for a large barrier $U$, with $U^*$ a characteristic energy (with $U$ and $U^*$ in units of temperature). 
If temperature is small enough, the typical time to overcome $U$ 
is given by the Arrhenius law, $\delta \sim t^* \exp[U/T]$, 
with $t^*$ a characteristic time. 
Changing variables we obtain 
$\psi(\delta) \sim \alpha {t^*}^{\alpha} \;\delta^{-(1+\alpha)}$, 
with $\alpha \sim T/U^*$.
Since 
clustering 
implies that not all EE have the same $U$ we will argue that 
the $\delta_i$ corresponds to
the special EE that act as CE epicenters. 
These EE may be associated to the ones allowing to 
escape from dominant configurations
~\cite{kolton_dep_zeroT_long}.
Two different  predictions for $U^*$ and thus for $\alpha$  
are found in the literature. 
In Ref.~\onlinecite{chen_marchetti_vinokur_1996}, 
it is assumed that $U^* \equiv T_d$. 
with $T_d$ 
From Table \ref{tabla} we obtain 
$\alpha=0.04$ for $T=293K$ and $\alpha=0.06$ for 
$T=323K$. Both results are in excelent agreenment 
with our data, which gives 
$\alpha \approx 0.05$.
In Ref.~\onlinecite{monthus_garel_2008} on the other hand, 
the characteristic energy is 
taken as the optimal nucleous barrier  
$U^* \equiv T_d(H_d/H)^{\mu}$, 
with $\mu=1/4$ for the one dimensional elastic string.
The exponent is thus again nonuniversal 
but now it is also field dependent, 
$\alpha \approx (T/T_d)(H/H_d)^{\mu}$.
From table \ref{tabla} we obtain 
$\alpha \approx 0.02$ both for the two temperatures 
and their corresponing fields.
This value is only slighlty below but is again 
of the order of $\alpha\approx 0.05$ we infer 
from our measurements.
Both predictions are in rough agreenment with the 
empirical $\alpha \approx 0.05$ we obtain from the
time-scaling of $S_{\tt WE}$ we observe in Fig.~\ref{fig:hist}(c)
for the two temperatures. It would be interesting to 
perform a more systematic study as a function of $T$ and $H$ 
to further test these theories. 
}
The previous observations 
lead us to argue that 
~{
WE give access not only to the CE area 
(at small $\Delta t$) 
but also to the waiting-time statistics (at larger $\Delta t$). 
As CE start at a 
seed EE, the $\delta_i$ must be controlled by their energy barrier distribution}
\footnote{Barrier distributions for dominant metastable states have been computed in Ref.~\onlinecite{kolton_dep_zeroT_long} 
for forces below and near $f_c$. For small systems a roughly exponential right tail 
followed by a cut-off can be appreciated.}.

\AKcom{\subsection{Event correlations}}

In order to further test the connection between WE and CE
we have also studied correlations from the spatio-temporal correlations of 
the registered positions $x_i$ of the $N$ measured WE epicentres. 
To do that we used the mean square distance 
$\langle \delta^2 x(\mathcal{T}) \rangle \equiv \sum_{i=1}^{N} [x_{i+n}-x_i]^2 /N$, 
which depends only on the temporal separation $\mathcal{T} = n t_0$.
For non-correlated WE epicentre sequences, $\langle \delta^2 x(\mathcal{T}) \rangle$ tends to a constant value 
$\mathcal{C} = (L_0+1)(L_0+2)/2$, where $L_0$ is approximately the lenght of the DW in units of the spatial discretization~\footnote{L. Foini, A. Rosso, private communication.}.
\begin{figure}[!ht]
\begin{center}
\includegraphics[width=0.9\columnwidth]{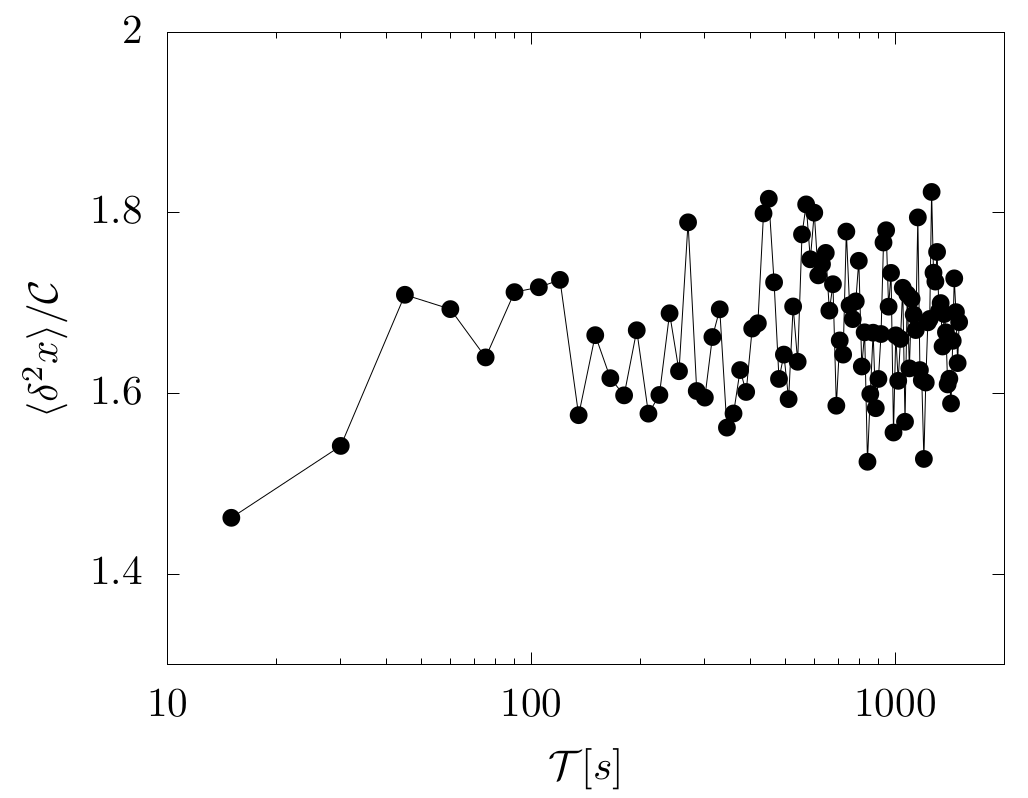}
\caption{Normalized mean square distance $\langle \delta^2 x(\mathcal{T}) \rangle/\mathcal{C}$ 
as a function of $\mathcal{T}$, measured at $T=$~RT and $H=46.1~\si{\oersted}$. 
}
\label{fig:correla-t0}
\end{center}
\end{figure}
Figure~\ref{fig:correla-t0} shows $\langle \delta^2 x(\mathcal{T}) \rangle/\mathcal{C}$
measured at $T=$~RT and $H=46.1~\si{\oersted}$. One can see that even for short $\mathcal{T}$ it becomes approximately constant as expected for uncorrelated events (note that $\langle \delta^2 x(\mathcal{T}) \rangle/\mathcal{C}>1$ for large $\mathcal{T}$ due to an underestimation of the length of DW).
We hence conclude that WE are very weakly correlated in sharp contrast with the predicted EE correlations in Ref.~\onlinecite{Ferrero2017} and consistent to what is predicted for CE and more generaly for depinning avalanches. 
This observation further confirms our identification of WE with single CE or with coalesced groups 
of them, for small or large $\Delta t$ respectively.

\subsection{Heuristic model for large WE}
\label{sec:discusions}
Summing up, our results are consistent with the predictions of Ref.\onlinecite{Ferrero2017} 
after identifying the small $\Delta t$ WE with the predicted CE. 
At large $\Delta t$ WE can not be single CE however, 
and deviations from the predicted properties for CE are expected. 
This is already apparent in Fig.~\ref{fig:pofl}(a) where 
large WE display a clearly different length to area aspect ratio
than the expected for CE. 
Moreover, Fig.~\ref{fig:pofl}(a) shows a clear crossover from the expected  
$S_i \sim L_i^{1+\zeta_d}$ CE behaviour to a different behaviour, 
rather well described by a new power-law, $S_i \sim L_i^{1.4}$. 
There is no theoretical predictions yet for this crossover so 
we propose here a simple, heuristic model.

\begin{figure}[!ht]
\begin{center}
\includegraphics[width=0.9\columnwidth]{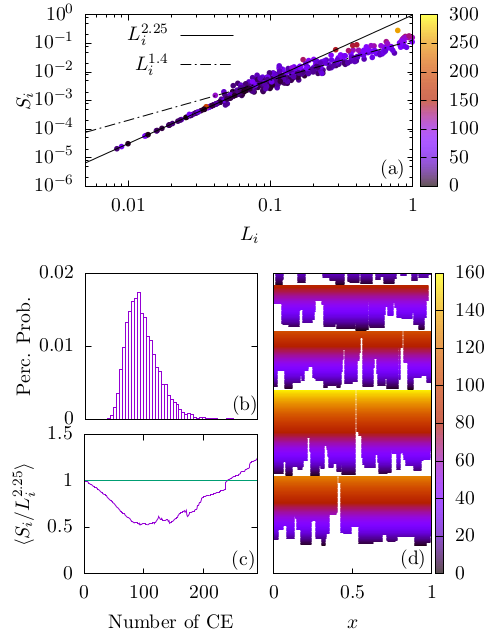}
\caption{A simple model for understanding the area 
vs major axis length of WEs. (a) Simulation results 
for the model (compare them experimental results in 
Fig.~\ref{fig:pofl}(a)). The solid line indicates the qEW depinning scaling, 
and the dashed dotted line the effective power law observed 
at large WE. The color bar indicates number of randomly deposited CEs.
Results are reported in arbitrary units.
(b) Probability distribution for percolation, as a function of 
the number of randomly deposited CE.
(c) Average aspect-ratio scaling versus number of CE. 
(d) From bottom to top, four actograms showing the growth 
of compact WE as we increase the number of deposited CE (as 
indicated by the color-bar). The actograms display 
segments centered at uniformely distributed epicenters, 
whose size equals the lateral extension of the growing
WE. When the system percolates, i.e. a single WE spans 
the system (set to unity), the actogram is reseted and 
a new deposition process starts.
}
\label{fig:model}
\end{center}
\end{figure}

A very simple model can explain the behaviour 
of $S_i$ vs $L_i$ observed experimentally, 
shown in Fig.~\ref{fig:pofl}(a).
The idea is to think WE as the compact objects formed 
by random deposition of simulated CE, 
with a lateral sizes $L_i$ sampled from 
$P_{\tt CE}(L)\sim L^{-\tau_L}$. We can also assume, 
for simplicity, that the corresponding areas 
satisfy a deterministic relation $S_i = L_i^{1+\zeta_d}$, 
assumption that leads automatically to $P_{\tt CE}(S)\sim S^{-\tau_{\tt CE}}$, 
with $\tau_L$, $\tau_{\tt CE}$ and $\zeta_d$ related by 
Eqs.(\ref{eq:tauS}) and (\ref{eq:tauL}). 
Both assumptions are reasonable approximations 
according to creep simulations~\cite{Ferrero2017}. 
To simulate this model we generated such events 
in the interval $[0,1]$, sequentially 
increasing the number of deposited CE.  
The process starts with one 
WE which equals the first deposited CE. Adding more CE may produce 
more WE (specially for small number of deposited CE) 
or can decrease their number due to the possible coalescence 
with an existing WE, if the new CE overlaps it. 
When a coalescence between a new CE and an existing WE takes place, 
the area of the resulting 
WE is the sum of the new CE area with the previous area of the 
WE, but the lenght of the new WE can either remain constant 
or increase at its left, rigth or both corners simultaneously.
In more rare cases the new CE can overlap more than one WE. 
The process finishes when a single WE spans the whole interval, 
i.e. when the deposited CE percolate the system. To make statistics 
over many sequences, at this point we reset the simulation 
and restart adding a first CE into a new actogram. 
Fig.~\ref{fig:model}(d) shows actograms corresponding to 
four runs.

To be concrete, for the simulations we use the values $\zeta_d=1.25$, $\tau=1.11$ and 
$\tau_L=\zeta_d$ corresponding to the 1d qEW 
depinning class. We sample the epicenter of each CE 
from a uniform distribution in the interval $[0,1]$ 
and its lateral size $l_i$ by
$l_i=0.05 [r(l_{\tt max}^{\tau_L+1}-l_{\tt min}^{\tau_L+1})+
l_{\tt min}^{\tau_L+1})]^{1/(\tau_L+1)}$ with 
$l_{\tt min}=10^{-5}$ and $l_{\tt max}=1.5$, and 
$r$ a different uniform random number in the interval $[0,1]$.
This produces a power-law distribution 
$P_{\tt CE}(l)\sim l^{-\tau_L}$ with a cut-off at $l_{\tt max}$. 
The area of such CE is simply $s_i=l_i^{1+\zeta_d}$.
Using different parameters yields qualitatively 
similar results.

We now discuss the results of the model.
In Fig.~\ref{fig:model}(a) 
we show that the model reproduces the main features
observed experimentally in Fig.~\ref{fig:pofl}(a). Small WE, below a characteristic 
crossver scale, display the $S_i \sim L_i^{2.25}$. 
This is natural, as most of the small WE 
are individual CE. At approximately $L^*\sim 0.07$ 
there is a crossover towards $S_i \sim L_i^{1.4}$ 
for large WE. Remarkably, this new exponent 
is indistinguishable from the one we obtain experimentally
(see Fig.~\ref{fig:pofl}(a)). We leave for a future study 
to understand the origin and possible universality 
of this new exponent. 
It is worth noting however, that the crossover 
may be associated to the lateral acceleration 
that occurs when the WE almost percolate the 
$[0,1]$ interval. In Fig.~\ref{fig:model}(b) 
we show the probability to percolate 
as a function of the number of deposited CE. 
Most of the points in Fig.~\ref{fig:model}(a) in the 
$S_i \sim L_i^{1.4}$ regime, belong to states with a 
high probability to percolate. It is also worth 
noting that finite size effects, due to the finiteness of the 
interval and the broad range of the CE lateral size distribution 
play an important role here.
In ~\ref{fig:model}(c) we show that the average 
anisotropic aspect-ratio $S_i/L_i^{2.25}$ is unity 
only for a small number of CE, but then decreases 
implying an accelerated lateral growth of WE, compared 
with the area growth. This effect may produce the 
crossover and the downward deviation appreciated 
in Fig.~\ref{fig:model}(a) and also experimentally 
shown in Fig.~\ref{fig:pofl}(a). For an even larger 
number of CE, the aspect-ratio increases as 
large WE tend to completely overlap with most of the new 
CE and thus increase their areas without modifying 
its lateral size. Those states are near to percolate 
but need a rare CE to overlap the voids between the 
few remaining WE.

The model presented has some unphysical features. In particular, 
the random deposition process implies, in the long time 
limit, a growing interface with a non-stationary width.
The model describes satisfactorily 
the crossover in $S_i$ vs $L_i$ observed in the experiments 
however, so the necessary surface relaxation effects 
or correlations that may make the width to saturate are 
not relevant for the regime we aim to describe.  
In addition, the number of deposited CE does not strictly represent 
time. Broadly distributed times between random depositions could 
be easily added however, in order to further test the picture 
suggested by the experiments. Particularly, to reproduce 
the area and lateral size distributions as a function of 
the window time $\Delta t$ found experimentally.

\section{Conclusions}
From our results the following picture emerges. 
Creep dynamics is driven by EE with 
a broad size distribution and a large size 
cut-off 
controlling 
the mean velocity.
The seed EE that trigger a cascade of extra EE 
are separated by broadly distributed waiting times. 
Repeated, this collective process of ignition and correlated growth 
produce independent CE statistically very similar to depinning 
avalanches, that may coalesce into larger compact objects. 
Hence, CE can be truly regarded as ``creep avalanches''.
{
The described picture}{, that drastically changes  
the naive view of creep motion as independent 
thermally nucleated displacements,}  
{is likely to appear not only in other magnetic films 
but in the creep regime of other disordered 
elastic systems in general. 
}

\begin{acknowledgments}
We thank E. Ferrero, A. Rosso, E. Jagla, G. Durin, T. Giamarchi, P. Le Doussal, L. Foini, 
V. Lecomte,  and J. L. Barrat for enlightening discussions. 
We thank D. Jordan for his help on digitizing DW.
This work was partly supported by grants PIP11220120100250CO/CONICET,
PICT2016-0069/FONCyT and UNCuyo 06/C490 and C017, from Argentina. 
The France-Argentina project ECOS-Sud No. A12E03 is also acknowledged.
\end{acknowledgments}

\appendix
\section{Samples \& Experimental Protocol}
\label{sec:samples}
Experiments were \AKcom{mainly} performed on a Pt/Co/Pt ultrathin ferromagnetic film, a prototypical system which has been the focus of many studies of domain wall motion~\cite{lemerle_domainwall_creep, metaxas_depinning_thermal_rounding, emori2012, Moon2013, Gorchon2014, Hrabec2014, Jeudy2016, Wells2017}. The studied sample was a Pt(4.5nm)/Co(0.7nm)/Pt(3.5nm) thin film, with the thickness of each layer indicated in parenthesis. The film was sputter grown at 300 K on etched Si/SiO$_2$ substrate. The magnetic response of this system to an external out-of-plane magnetic field is characterized by a square magnetic cycle with a well defined remnant magnetization, typical of systems with perpendicular magnetic anisotropy~\cite{metaxas_depinning_thermal_rounding}.
%

Polar magneto-optical Kerr effect (PMOKE) microscopy has been used to image magnetic domains after applying magnetic field pulses perpendicular to the film plane. After fully magnetizing the sample in one direction, a short pulse in the opposite direction and of intensity $H =130$~Oe was first applied in order to nucleate a seed magnetic domain. Then, a second pulse of duration $t$ and intensity $H$ was applied to favour the growth of the initial magnetic domain. DW velocity was then computed as the ratio between the linear advance of the DW linear advance $\Delta x$ and the pulse duration, $v=\Delta x/t$. 
Experiments were performed at $T=$~RT (room temperature) and $T=50~\si{\celsius}$.
To measure velocities between $10^{-9}$ m/s and $10^{-1}$ m/s pulses of different amplitude and duration were used. In all the cases, the total number of pulses was 15 or more and the rise time of the pulses was more than one order of magnitude faster than the pulse duration. The shortest pulse was 1 ms and the largest one 1800 s. 
Due to the spatial resolution of our microscope, for velocities smaller than $10^{-8}$ m/s we observed that there was no difference in the observed DW velocity if the magnetic field was ON or OFF during the image acquisition. In all the cases given the illumination condition, the used shutter time of the camera was 200 ms. 

\AKcom{Although we mainly report results for one region of a specific sample, we have made similar measurements in other regions of the same sample and also in a Pt(6nm)/[Co(0.2nm)/ Ni(0.6nm)]$_3$/Al(5nm) sample, where the numbers in parenthesis stand for thickness and the ferromagnetic layer consists in a stack of three Co(0.2nm)/ Ni(0.6nm) bilayers (for more information about these samples and their domain wall dynamics see Refs.~\onlinecite{Rojas2016, Caballero2017}).
In both cases the results are consistent with the main universal results reported for the specific region of the Pt/Co/Pt sample in Section~\ref{sec:results}.}

\begin{figure}[!ht]
\begin{center}
\includegraphics[width=0.9\columnwidth]{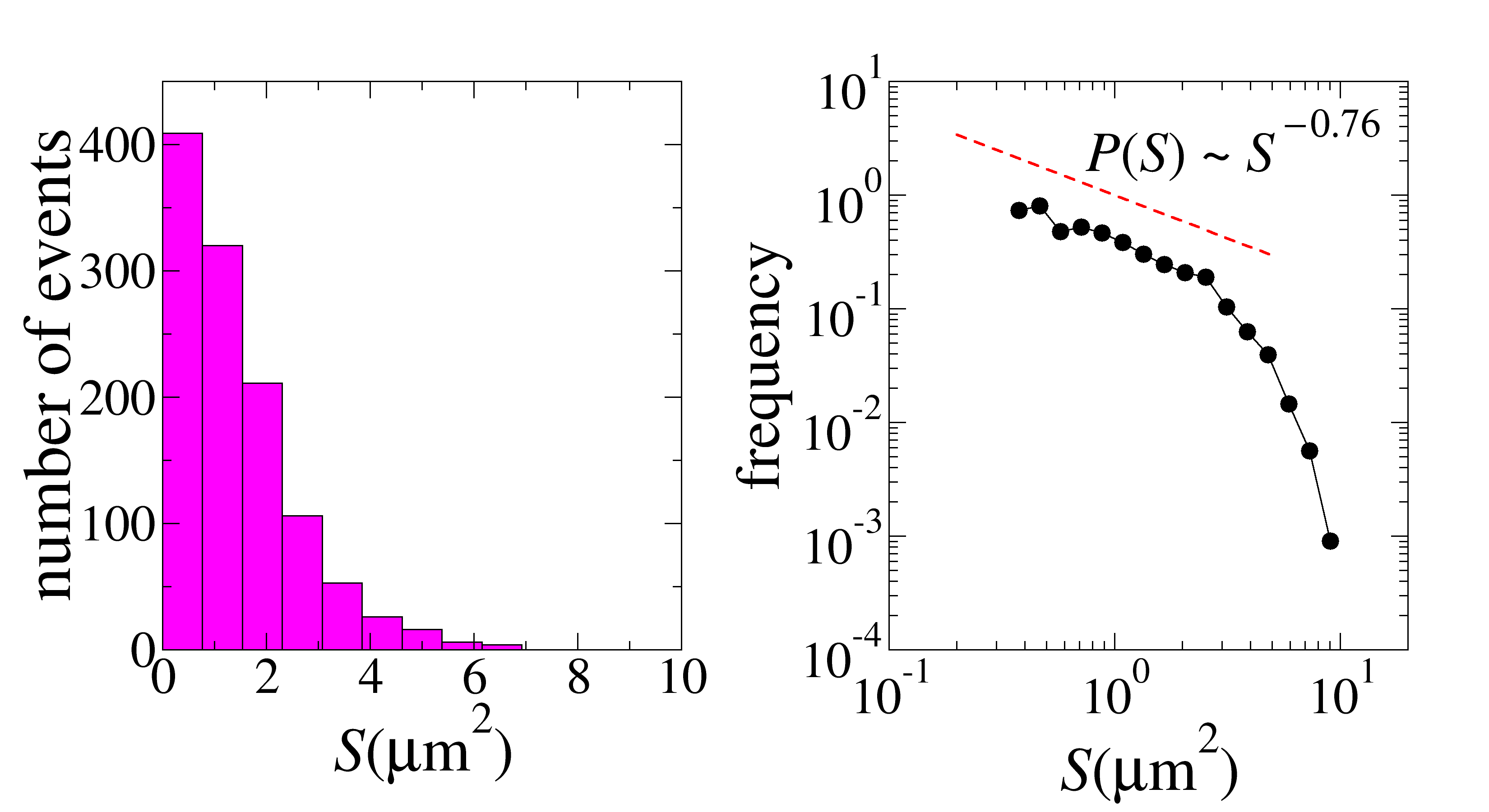}
\caption{Histograms for the $N_{\tt tot}(t_0,t)=1151$ WE shown in Fig.~\ref{fig:creep-plot}(c), obtained by comparing consecutive images taken every $\Delta t= t_0 =15~\si{\second}$ at $T=$~RT and $H=46.1~\si{\oersted}$.
The histogram with uniform binning is presented in the left panel, while in the right panel a logarithmic binning is used for same WE. Power-law behaviour at small sizes can be described with $P_{\tt WE}(S) \sim S^{-\tau_{\tt WE}}$ with the distribution exponent $\tau_{\tt WE} \approx 0.76 \pm 0.1$. 
}
\label{fig:hist-t0}
\end{center}
\end{figure}
Magnetization reversal events were previously obtained in irradiated Pt/Co/Pt samples~\cite{Repain2004}, identifying between 30 and 50 events depending on field values.
In the present work, as we previously anticipated, we were able to obtain a large amount of WE. This represents a quantitative progress in view of the fact that this allows us to perform a deeper statistical description of the data. Figure~\ref{fig:hist-t0} shows the obtained histogram of the 1151 WE areas shown in Fig.~\ref{fig:creep-plot}(c) by comparing consecutive images taken every $t_0=15~\si{\second}$.
Since we are seeking power-law like distributions and their effective exponents, it is convenient to use logarithmic binning. 
In the right panel of Fig.~\ref{fig:hist-t0} we use the same WE used in the left panel to build a new histogram. 
In this case, dividing the number of events per interval by the width of the interval, the probability 
distribution is obtained. 
Fig.~\ref{fig:hist-t0} shows a power-law signature at small size values with a cut-off around 3~$\si{\micro \meter}^2$. The distribution is of the form $P_{\tt WE}(S) = S^{-\tau_{\tt WE}} G_{\tt WE}(S/S_{\tt WE})$, where $\tau_{\tt WE}$ is the power-law exponent and $S_{\tt WE}$ the cut-off value such that the function $G_{\tt WE}(x)$ rapidly decays for $x \gtrsim 1$. 

\AKcom{
For a proper comparison 
with theoretical predictions we discarded in the statistical analysis events touching the borders of the region of interest, i.e. the observation region, otherwise their area would be underestimated.
This may affect however the tails of the size distribution, corresponding 
to large events, with a lateral size of the order or larger 
than the lateral size $L$ of the observation region. 
For the range of time windows $\Delta t$ we consider WE of lateral size $L$
are extremely rare however. 
Indeed we observe a $\Delta t$-dependent but clearly $L$-independent cut-off in our distributions, 
growing as $\Delta t^{1/2}$ (with a temperature dependent prefactor). Such $\Delta t$ 
dependence is used to estimate the waiting time distribution exponent 
for cluster “ignition events”. On the other hand, the power-law decay 
effective exponent of WE, which is also central to our analysis and for the 
comparison with theory is not sensible to the tails. This justifies our event detection 
protocol.
}

\AKcom{
To evidence the robustness of our results, in Fig.~\ref{fig:hist-otherregionssamples} 
we show results for the event area distribution measurements done in a different region of the same Pt(4.5nm)/Co(0.7nm)/Pt(3.5nm) 
sample (upper panel), and for a Pt(6nm)/[Co(0.2nm)/ Ni(0.6nm)]$_3$/Al(5nm) sample (lower panel).
As can be appreciated, not only the effective power-law decay exponent is similar, 
but also the time dependence is qualitatively similar to the one of Fig.~\ref{fig:hist} reported in 
in Sec.\ref{sec:results} for the sample and region we have chosen for most of our analysis. 
}
\begin{figure}[!ht]
\begin{center}
\includegraphics[width=0.7\columnwidth]{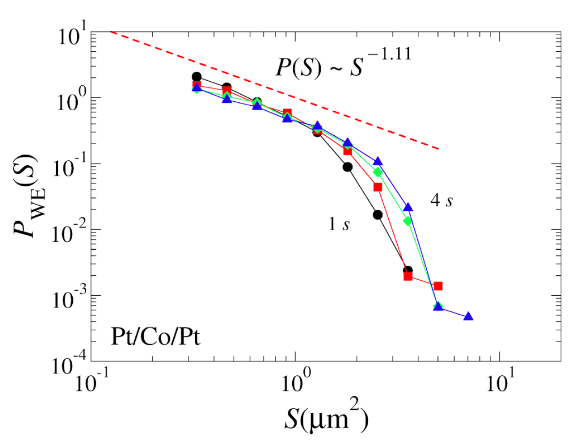}
\includegraphics[width=0.7\columnwidth]{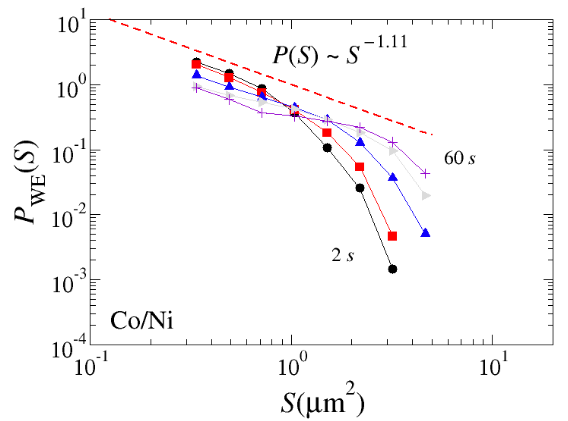}
\caption{
\AKcom{WE area histograms obtained from a different region (upper panel) 
of the same Pt(4.5nm)/Co(0.7nm)/Pt(3.5nm) sample used to report most our results in the main text, 
and obtained for a different Pt(6nm)/[Co(0.2nm)/ Ni(0.6nm)]$_3$/Al(5nm) sample (lower panel).
}
}
\label{fig:hist-otherregionssamples}
\end{center}
\end{figure}

\section{Domain wall roughness}
\label{sec:roughnessmethod}
In order to have a more direct estimate of the roughness exponent 
of our DW, and check consistency with our interpretaion of the WE statistics, 
we have computed the single-value displacement field $u_L(x)$ 
of segments of given sizes $L$ partitioning a larger DW configuration.
Here we discuss our practical method.
The displacements $u_L(x)$ for each segment are measured with 
respect to the straight line fitting each segment. 
This straigth line is also used as the $x$-axis to parametrize 
the displacement field. 
Such an approach is justified by taking into account 
that the theoretical description of a directed driven interface 
assumes that the interface is flat in average 
in the direction perpendicular to the motion. The field 
on magnetic DW on the other hand act as a pressure, allways normal 
to the DW. 
Having $u_L(x)$ for segments of different length $L$, we can now 
compute their global squared width 
$W^2(L) \equiv \overline{u^2_L(x)}-\overline{u_L(x)}^2$. 
If the interface is self-affine, we expect $W^2 \sim L^{2\zeta}$ 
with a well defined roughness exponent $\zeta$.
We have tested this methodology numerically on 
large discretized interfaces of size $L_0$, with 
displacement field $U(x)$, where $x=0,1,...,L_0-1$,
with different precise values of $\zeta$. We do so  
by superimposing Fourier modes $U(x)=\sum_q U_q e^{iqx}$ 
with $q=2\pi n/L_0$ ($n=0,...,L_0-1$), $U_q$ complex hermitian 
gaussian amplitudes of zero mean, $\langle U_q \rangle=0$, 
and variance $\langle |U_q|^2 \rangle \sim 1/q^{1+2\zeta}$ 
(also known as ``self-affine gaussian signals''\cite{gaussiansignals2005}). 
This construction assures that the signal $U(x)$ is periodic, $U(x)=U(x+L_0)$, 
and self-affine with identical spectral and global exponent $\zeta$. 
Finally, we compute $W^2(L)$ for these interfaces, by the partition 
procedure previously described. 
 \begin{figure}[!ht]
\begin{center}
\includegraphics[width=0.9\columnwidth]{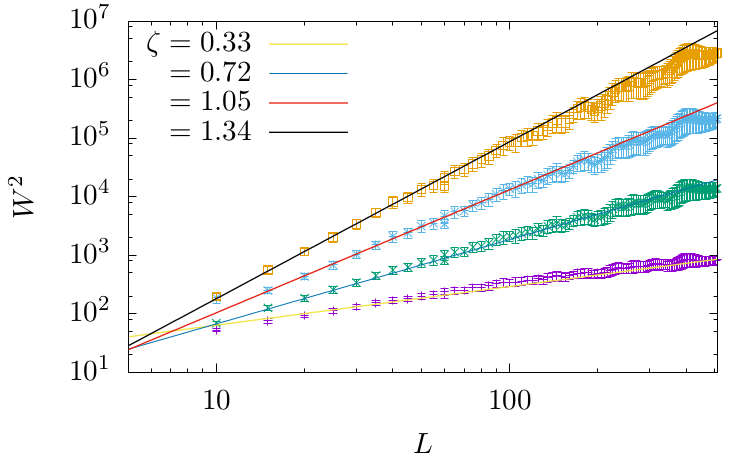}
\caption{Numerical test for the 
practical implementation used to compute $W^2$ for a DW. 
We compute $W^2$ for numerically generated self-affine gaussian 
signals of size $L_0=1024$ for several precise values of 
$\zeta$. We average over $10$ samples for each $\zeta$. 
The solid lines show agreenment with the expected 
$W^2\sim L^{2\zeta}$. The method allows to measure 
super-rough cases $\zeta>1$.}
\label{fig:check}
\end{center}
\end{figure}
In Fig.~\ref{fig:check} we compare $W^2(L)$ vs $L$ with the corresponding 
scalings for each $\zeta$, averaged over $10$ uncorrelated numerically sampled 
configurations $U(x)$. 
A good agreenment is allways obtained if the fit does not include values
of $L$ larger than a fixed fraction of the order of the total size $L_0$,
so to have a large number of segments and to reduce 
boundary effects.
It is worth noting that our method also 
allows to accurately measure values $\zeta>1$, corresponding 
to super-rough interfaces. This is an advantage over
the displacement correlator function 
$B(x) \equiv \int_0^{L_0-x} dx_0\;\overline{[u(x+x_0)-u(x_0)]^2}/(L_0-x)$ which 
gives the correct global $\zeta$, $B(x) \sim x^{2\zeta}$, only if $\zeta<1$, otherwise
it saturates to $\zeta=1$.

\bibliography{tfinita5,paredom,dmi,universal}

\end{document}